\documentclass[a4paper, 12pt, fleqn]{article} 
\pdfoutput=1 
\usepackage{myarticle}

\title{\fontsize{17pt}{1cm}\selectfont{}Low Government Performance and Uncivil Political Posts on Social Media: Evidence from the COVID-19 Crisis in the US $^*$}
\author{Kohei Nishi $^\dagger$}
\date{}
\begin{document}
\maketitle
\vspace{-8mm}

\begin{center}\textbf{Abstract}\\\end{center}
\noindent
{\fontsize{10.5pt}{0.35cm}\selectfont
Political expression through social media has already taken root as a form of political participation. Meanwhile, democracy seems to be facing an epidemic of incivility on social media platforms. With this background, online political incivility has recently become a growing concern in the field of political communication studies. However, it is less clear how a government's performance is linked with people's uncivil political expression on social media; investigating the existence of performance evaluation behavior through social media expression seems to be important, as it is a new form of non-institutionalized political participation. To fill this gap in the literature, the present study hypothesizes that when government performance worsens, people become frustrated and send uncivil messages to the government via social media. To test this hypothesis, the present study collected over 8 million posts on X/Twitter directed at US state governors and classified them as uncivil or not, using a neural network-based machine learning method, and examined the impact of worsening state-level COVID-19 cases on the number of uncivil posts directed at state governors. The results of the statistical analyses showed that increases in state-level COVID-19 cases led to a significantly higher number of uncivil posts against state governors. Finally, the present study discusses the implications of the findings from two perspectives: non-institutionalized political participation and the importance of elections in democracies.

\vspace{13mm}
\noindent\rule{2in}{0.5pt}

\noindent
$^*$ The first version of this working paper was uploaded on July 21, 2021, and this version on January 19, 2024. Note that it has not yet passed peer review.

\noindent
$^\dagger$ Research Fellow of Japan Society for the Promotion of Science and Ph.D. Student at the Division of Law and Political Science, the Graduate School of Law, Kobe University, Japan

}

\newpage
\section*{Introduction}
Nowadays, many people use social media platforms, such as X/Twitter and Facebook, to express their political opinions. With this background, political expression through social media has already become a form of political participation (Theocharis, 2015)\nocite{Theocharis2015}, and related research has been rapidly advancing. Scholars have found, for example, that social media are prone to selective exposure and the echo chamber phenomenon (Bakshy et al., 2015; Cinelli et al., 2021; Mosleh et al., 2021)\nocite{Bakshy2015, Cinelli2021, Mosleh2021}. It has also been found that political engagement on social media enhances offline political participation (Bode, 2017; Conroy et al., 2012; Dimitrova et al., 2011; Holt et al., 2013)\nocite{Bode2017, Conroy2012, Dimitrova2011, Holt2013}. These findings suggest the importance of further research in this area for a better understanding of contemporary politics.

Since the outbreak of the COVID-19 pandemic, people have had conversations on social media about the pandemic (Xiong et al., 2021; Xue et al., 2020)\nocite{Xiong2021, Xue2020}. According to previous studies, communication about COVID-19 on social media is politicized (Jiang et al., 2020)\nocite{Jiang2020}, and echo chambers can be observed therein (Jiang et al., 2021)\nocite{Jiang2021}. In addition, conspiracy theories about COVID-19 and hate speech against Asians are prevalent on social media (Ahmed et al., 2020; He et al., 2021)\nocite{Ahmed2020, He2021}. Thus, research on social media communication is also important in the context of the COVID-19 pandemic.

While previous studies have investigated the nature of political communication on social media from a variety of perspectives, few have examined how the worsening of government performance is linked with people's uncivil political expression on social media. However, investigating how people manage to monitor and control their governments through social media expression, a form of non-institutionalized political participation (Michalski et al., 2021)\nocite{Michalski2021}, is important from a political science perspective. The present study aims to fill this gap in the literature.

It hypothesizes that when government performance worsens, people feel frustrated and angry and thus post uncivil messages on social media to berate the government. The present study tested this hypothesis in the context of the COVID-19 pandemic by collecting a large number of X/Twitter posts directed at US state governors and classifying them as uncivil or not using a neural network machine learning method. Using these data, it examined the impact of worsening state-level COVID-19 indicator on the number of uncivil posts directed at state governors. The results showed that the number of state-level COVID-19 cases positively affected the number of uncivil posts directed at state governors.
This serves as evidence that the worsening of government performance is linked with people's uncivil political expression on social media.

\section*{Political Communication on Social Media}
Social media seems to have played a significant role in real-world changes, such as the outbreak of the Arab Spring (Howard et al., 2015; Waechter, 2019)\nocite{Howard2015, Waechter2019} and the electoral victory of Donald Trump in the 2016 US presidential election (Enli, 2017; Francia, 2017)\nocite{Enli2017, Francia2017}. With this background, many scholars have studied political communication on social media from a variety of perspectives.

Some scholars have focused on what motivates people to express their political opinions on social media. For example, Bekafigo and McBride (2013)\nocite{Bekafigo2013} have found that people with stronger partisanship levels and higher political engagement are more likely to post about politics on X/Twitter. In terms of elections, it has been found that many people use social media during elections to persuade others to vote for the party that they support (Hosch-Dayican et al., 2016)\nocite{Hosch-Dayican2016}. In terms of social movement, it has been revealed that people participate in social media protests, such as the \#MeToo movement, to change society (Mendes et al., 2018)\nocite{Mendes2018}. Overall, these findings seem to suggest that people engage in political expression on social media to change other people's opinions and social conditions.

In addition, online incivility has recently become a growing concern in the field of political communication studies (e.g., Borah, 2013; Jamieson et al., 2017; Kenski et al., 2017; Muddiman \& Stroud, 2017; Papacharissi, 2004; Sobieraj \& Berry, 2011; Theocharis et al., 2016)\nocite{Borah2013, Jamieson2017, Kenski2017, Muddiman2017, Papacharissi2004, Sobieraj2011, Theocharis2016}. Previous studies have suggested that the prevalence of online political incivility has various consequences in democracies. For example, people's exposure to incivility in discussions can lead to negative feelings toward the  discussion partner (Hwang et al., 2018; Kim \& Kim, 2019)\nocite{Hwang2018, Kim2019} and lower perceptions of the rationality of the opponent's argument (Popan et al., 2019)\nocite{Popan2019}. These findings suggest that online incivility prevents democratic society from reaching a consensus. People's frequent exposure to online incivility has also been found to be negatively related to their levels of online and offline political participation (Yamamoto et al., 2020)\nocite{Yamamoto2020}, suggesting that online incivility decreases people's political participation. Thus, because online incivility affects various aspects of democratic politics, understanding the factors that contribute to online incivility is crucial to maintaining democracy. In this context, a growing body of literature has investigated the factors shaping online political incivility (e.g., Coe et al., 2014; Gervais, 2017; Rains et al., 2017; Rowe, 2015; Vargo \& Hopp, 2017)\nocite{Coe2014, Gervais2017, Rains2017, Rowe2015, Vargo2017}.

\section*{Government Performance}
To understand political communication, it is also important to comprehend how people perceive their government, as this is a key actor in the political process. Political scientists have long investigated the relationships between people's evaluations of their government's performance and their voting behaviors (e.g., de Vries \& Giger, 2014; Ecker et al., 2016; Fiorina, 1978; Fournier et al., 2003; Hobolt et al., 2013; Kinder \& Kiewiet, 1979)\nocite{DeVries2014, Ecker2016, Fiorina1978, Fournier2003, Hobolt2013, Kinder1979}, showing that when government performance (such as the economy) is good, people vote for the government's party to reward it; however, when the performance is bad, people vote for the opposition to punish the incumbent government. Such voting patterns are called retrospective voting.

While the concept of government performance has been frequently used to explain people's voting behavior, few studies have examined whether low government performance leads to people's uncivil political expression on social media. However, it is crucial to investigate the relationships between government performance and online political incivility, as political expression on social media is a new form of non-institutionalized participation (Michalski et al., 2021)\nocite{Michalski2021}, which is theoretically important as a tool for expressing political grievances and challenging authorities (Melo \& Stockemer, 2012; Norris, 2002)\nocite{Melo2012, Norris2002}. 

To address this issue, the present study argues that people's evaluations of government performance lead to political expression on social media. More specifically, the present study hypothesizes that the worsening of government performance makes people frustrated and angry with the government and thus leads them to engage in uncivil political expression against the government on social media.

Evidence from previous studies indirectly supports this argument. It has been shown that US presidential approval ratings in public opinion polls correlate with the public sentiment expressed on X/Twitter (O'Connor et al., 2010)\nocite{OConnor2010}. This suggests that people's evaluations of their government lead to their political expression on social media. However, it is unclear whether the correlation shown by O'Connor et al., (2010)\nocite{OConnor2010} is in response to government performance. In addition, previous studies in social psychology have shown that frustration states lead to aggressive behavior toward the person causing the frustration (Berkowitz, 1989; Dill \& Anderson, 1995; see also Breuer \& Elson, 2017)\nocite{Berkowitz1989, Breuer2017, Dill1995}, which suggests that people's political frustration may be a cause of online political incivility. The combination of these findings strengthens the present study's argument.

The above mechanism should work as follows for the COVID-19 pandemic age. Throughout the pandemic, the number of COVID-19 cases has been considered an important indicator of government performance because people expect politicians to control the situation through policies related to lockdowns, masks, vaccines, and so on. Based on this assumption, when the number of COVID-19 cases increases (i.e., the COVID-19 indicator worsens), people would be expected to become frustrated and angry with the government, which is viewed as responsible for the worsening situation, and thus send uncivil messages to it on social media. Accordingly, the present study introduces the following hypothesis:

\vskip\baselineskip
\noindent
\textbf{Hypothesis}: An increase in COVID-19 cases leads to a higher number of uncivil messages on social media directed at governments.

\section*{Methods}
To test this hypothesis, the present study constructed a state-level time series dataset that recorded certain COVID-19-related indicators and the numbers of uncivil posts on X/Twitter directed at governors. Using the dataset, fixed effects regression models were estimated with ordinary least squares (OLS).

\subsection*{Collecting Posts}
The present study collected replies or mention posts directed at the X/Twitter accounts of US state governors that were posted between April 1, 2020, and March 31, 2021, using X/Twitter API for Academic Research.\footnote{The collection of posts was mainly conducted on May 21 and 22, 2021. For Rhode Island's new governor, the collection was conducted on April 24, 2023.} In the case of Montana, Utah, and Rhode Island, where the governors changed in the middle of the data-collection period, the target accounts of the collection were switched on the inauguration day of the new governors. Through these procedures, 8,045,894 posts were eventually collected.

\subsection*{Classifying the Posts Using Machine Learning}
The collected posts were classified as uncivil or not using BERT (Devlin et al., 2019)\nocite{Devlin2019}, a neural network machine learning method for language processing.

A supervised dataset for training and testing the classification model was constructed via Lucid Marketplace as follows:\footnote{This was conducted after review and approval by the research ethics committee (IRB) of the Graduate School of Law, Kobe University (ID: 030013).}
First, 2,000 posts were randomly extracted from the collected posts.\footnote{Posts that were not public at the time of the random extraction were excluded from the target of the random extraction. They were not public at the time presumably because they were deleted, the accounts were deleted, or the accounts were changed to private after they were collected. Because non-public posts cannot be displayed using X's embed post function (a function that was used to present posts to respondents), such posts were excluded from the target of the random extraction.}
Then, US respondents recruited via Lucid were presented with the posts using X's embed post function (\url{https://help.twitter.com/en/using-x/how-to-embed-a-post}) and classified them as uncivil or civil in accordance with the following definition of incivility: ``a disrespectful or insulting expression that attacks an individual or group.''
The respondents classified 20 posts per person per participation.
Each post was classified by 16.8 respondents on average.
Posts classified as uncivil by a majority of respondents were labeled as uncivil, while those classified as civil by a majority were labeled as civil in the supervised dataset.\footnote{300 US citizens between the ages of 18 and 70 were recruited, which was conducted over five rounds. A quota sampling approach based on age and gender was employed to obtain a sample that closely resembled the composition of the US population. For a technical reason, the number of participants exceeded the target numbers for some strata. While the author initially planned to use data excluding excess responses, due to a coding error of data processing, machine learning was conducted using data including them. Thus, the present study reports the analysis with data including excess responses. Using all responses including excess responses has the advantage of making the voting results closer to the true values in terms of the law of large numbers, but has the disadvantage of biasing the composition of the respondents. Conversely, excluding the responses of those in excess of the target numbers of strata has the opposite advantage and disadvantage. Note that the responses of those who failed an attention check question and those who answered ``A text is not displayed'' were excluded when the supervised dataset was constructed.}

The percentage of inter-coder agreement was 62.4\%.\footnote{To calculate the percentage of inter-coder agreement, the percentage of two-coder agreements in all two-coder combinations was computed for each post, and then the percentages for all posts were averaged. The responses of those who failed in an attention check question and those who answered ``A text is not displayed'' were excluded when the percentage of inter-coder agreement was calculated.} While this is a less favorable outcome, this can be justified by considering that the value is higher than 50\%, which would be the value expected if the coders were to annotate posts randomly. It is also important to note that, given that previous research has shown that annotator's judgments in toxicity annotation tasks are associated with an annotator's race and political attitudes (Sap et al., 2021)\nocite{Sap2021}, it may be generally difficult to achieve high agreement rates on political incivility annotation tasks.

With the supervised dataset, the present study trained the neural network machine learning model based on BERT to construct a classifier. 
The use of BERT to detect hate speech or incivility on social media is an accepted approach that has been employed by several previous studies (e.g., Mozafari et al., 2019; Trifiro et al., 2021)\nocite{Mozafari2019, Trifiro2021}.
In the present study, BERTweet-base (Nguyen et al., 2020)\nocite{Nguyen2020} was used as a pre-trained BERT model, and the outputs of the pre-trained layer were passed to a dropout layer, a linear layer, and then a sigmoid function. The process was conducted five times with different random seeds, and then the sigmoid outputs of the five times were averaged (random seed averaging; see Morzhov, 2021)\nocite{Morzhov2021}.\footnote{The training of the classification model was conducted based on PyTorch (Paszke et al., 2019)\nocite{paszke2019pytorch} and Transformers (Wolf et al., 2020)\nocite{Wolf2020} in Python.} Table \ref{classification_performance} shows the classifier's performance, calculated via five-fold cross-validation.
The scores indicate the successful detection of incivility with high performance, compared to previous studies.\footnote{The present study has a better F1-score than previous studies that have used machine learning methods to detect political incivility. For example,  Trifiro et al. (2021)\nocite{Trifiro2021} and Theocharis et al. (2020)\nocite{Theocharis2020} obtained F1-scores of .65 and .66, respectively. Note that the F1-score of Theocharis et al. (2020)\nocite{Theocharis2020} here was obtained based on the reported precision and recall scores.}

For additional validation of the classification performance, the top 50 frequent words were obtained for each of the posts classified as civil and uncivil.\footnote{To obtain the frequent words, all words in the collected posts were made lowercase, English stop words listed in Stopwords ISO (Diaz, 2020)\nocite{stopwordsiso} were excluded from them, and the words were lemmatized using WordNetLemmatizer in NLTK (Bird \& Loper, 2004)\nocite{bird-loper-2004-nltk} in Python. The top 50 frequent words were then obtained for each category, and letters that should be capitalized were manually capitalized (e.g., COVID).} As shown in Table \ref{frequent_words}, the top 50 words in posts classified as civil (the left column) do not include typical uncivil words, while the top 50 words in posts classified as uncivil (the right column) include some typical uncivil words (words in red and bold). This serves as an additional validation of the classification performance.

\begin{table}[H]
 \tablesize\centering
 \caption{Classification Performance}
 \label{classification_performance}
  \begin{tabularx}{95mm}{lRRRR}
   \hline
    & Accuracy & Precision & Recall & F1-score \\
   \hline
   Score & .838 & .654 & .747 & .697 \\
   \hline
  \end{tabularx}
\end{table}

\begin{table}[H]
    \tablesize\centering
    \caption{The Top 50 Frequent Words for Each Category}
    \label{frequent_words}
    \begin{tabularx}{170mm}{L|L}
        \hline
        The Top 50 Words in Posts Classified as Civil & The Top 50 Words in Posts Classified as Uncivil\\
        \hline
        people, mask, COVID, governor, time, school, day, Trump, job, business, life, death, vaccine, care, virus, vote, week, wear, family, stay, health, unemployment, mandate, money, kid, safe, NY, county, president, Texas, nursing, leadership, month, wearing, science, law, child, plan, home, Florida, pandemic, love, pay, public, country, election, Cuomo, citizen, testing, police & people, Trump, mask, governor, COVID, \textcolor{red}{\textbf{fuck}}, \textcolor{red}{\textbf{idiot}}, \textcolor{red}{\textbf{shit}}, death, \textcolor{red}{\textbf{ass}}, time, life, \textcolor{red}{\textbf{fucking}}, \textcolor{red}{\textbf{stupid}}, vote, job, resign, care, nursing, business, shut, virus, killing, lie, day, killed, home, Democrat, \textcolor{red}{\textbf{moron}}, kill, liar, money, Cuomo, sick, citizen, office, president, hope, election, country, American, Republican, NY, lying, Texas, wear, hand, disgrace, power, worst\\
        \hline
    \end{tabularx}
\end{table}

Then, the present study used this classifier to classify the collected posts as uncivil or not. As a result, 2,172,839 out of 8,045,894 posts (27.0\%) were classified as uncivil. This percentage of incivility is slightly higher than the values reported by previous studies (Coe et al., 2014; Theocharis et al., 2020; Trifiro et al., 2021)\nocite{Coe2014, Theocharis2020, Trifiro2021}.

\subsection*{Counting Uncivil Posts}
To count the daily numbers of uncivil posts, it was necessary to identify the date and time when they were posted. Because the date and time of the posts were originally recorded in Coordinated Universal Time (UTC), the present study converted them from UTC to the standard time of each state. In the case of states with multiple standard time zones, the standard time of the state capital was used. In addition, in the case of a daylight saving time period, the time was converted to daylight saving time.

Based on the identified date and time, the daily numbers of uncivil posts were counted for each state. As a rule, uncivil posts created by users who had sent uncivil posts to multiple state governors were excluded from the count. When a user had sent uncivil posts to several state governors, it meant that the user had sent the uncivil post to governors in states other than their own home state. However, the present study's theory posits that people send uncivil posts as a result of evaluating their own government's performance. As sending uncivil posts to multiple state governors lies outside the scope of the present study, such uncivil posts were excluded from the counts.

\subsection*{Constructing a Dataset}
Combining the above data with other data from several sources, the present study constructed a state-level time series dataset. The dataset comprised 18,250 observations (50 states $\times$ 365 days) and included state-level daily numbers of uncivil posts directed at state governors, posts posted by state governors, COVID-19 cases, PCR tests, and dummy indicators for the presence of lockdown and mask policies. The numbers of posts posted by state governors were retrieved via X/Twitter API. The numbers of COVID-19 cases and PCR tests were retrieved from the ``COVID-19 Diagnostic Laboratory Testing (PCR Testing) Time Series'' (U.S. Department of Health \& Human Services [HHS], 2023a, 2023b)\nocite{USDHHS2023,USDHHS2023_archive}.\footnote{The present study used the numbers of positive cases from PCR tests as an indicator of the numbers of COVID-19 cases. This dataset has several versions, which are listed by the archive repository (HHS, 2023b)\nocite{USDHHS2023_archive}. The present study used the version that was uploaded on May 30, 2023 (\url{https://us-dhhs-aa.s3.us-east-2.amazonaws.com/j8mb-icvb_2023-05-30T12-05-12.csv}).}
Indicators for lockdown and mask policies were retrieved from the ``Oxford COVID-19 Government Response Tracker'' (Hale et al., 2021, 2023)\nocite{Hale2021paper, Hale2023data}.\footnote{The present study used the file ``OxCGRT\_compact\_subnational\_v1.csv`` (retrieved July 27, 2023, from \url{https://github.com/OxCGRT/covid-policy-dataset/blob/main/data/OxCGRT\_compact\_subnational\_v1.csv}).
Regarding lockdowns, the dataset included the variable ``C6M\_Stay.at.home.requirements'', which was rated on an ordinal scale ranging from 0 to 3.
The present study re-coded this variable to take the value of 1 if the original value is 2 or 3 and 0 otherwise, to create a dummy variable for lockdown.
Regarding mask requirements, the dataset included the variable ``H6M\_Facial.Coverings'', which was scored on an ordinal scale ranging from 0 to 4.
The present study re-coded this variable to take the value of 1 if the original value is 3 or 4 and 0 otherwise, to create a dummy variable for strong mask requirement.
}
Table \ref{descriptive} presents the descriptive statistics for the dataset of the present study.

\subsection*{Estimation Strategies}
Using the dataset, two-way fixed effects regression models were estimated. More specifically, the following four models were estimated:

\begin{equation}
\tag{1}
\begin{split}
\ln(\textit{UncivilPosts}_{st} + 1) =\ & \beta_1\ln(\textit{COVID}_{st-1} + 1) + \beta_2\ln(\textit{PCR}_{st-1} + 1)\\
& + \beta_3\ln(\textit{GovernorPosts}_{st} + 1) + \beta_4\textit{Lockdown}_{st} + \beta_5\textit{Mask}_{st}\\
& + \textit{StateFE}_s + \textit{DayFE}_t + \epsilon_{st}
\end{split}
\end{equation}

\begin{equation}
\tag{2}
\begin{split}
\ln(\textit{UncivilPosts}_{st} + 1) =\ & \beta_1\ln(\textit{COVID}_{st-1} + 1) + \beta_2\ln(\textit{PCR}_{st-1} + 1)\\
& + \beta_4\textit{Lockdown}_{st} + \beta_5\textit{Mask}_{st}\\
& + \textit{StateFE}_s + \textit{DayFE}_t + \epsilon_{st}
\end{split}
\end{equation}

\begin{equation}
\tag{3}
\begin{split}
\ln(\textit{UncivilPosts}_{st} + 1) =\ & \beta_1\ln(\textit{COVID}_{st-1} + 1) + \beta_2\ln(\textit{PCR}_{st-1} + 1)\\
& + \beta_3\ln(\textit{GovernorPosts}_{st} + 1)\\
& + \textit{StateFE}_s + \textit{DayFE}_t + \epsilon_{st}
\end{split}
\end{equation}

\begin{equation}
\tag{4}
\begin{split}
\ln(\textit{UncivilPosts}_{st} + 1) =\ & \beta_1\ln(\textit{COVID}_{st-1} + 1) + \beta_2\ln(\textit{PCR}_{st-1} + 1)\\
& + \beta_3\ln(\textit{GovernorPosts}_{st} + 1) + \beta_4\textit{Lockdown}_{st} + \beta_5\textit{Mask}_{st}\\
& + \textit{StateFE}_s + \textit{DayFE}_t + \textit{StateFE}_s \times \textit{Trend}_{t} + \epsilon_{st}
\end{split}
\hspace{-10mm}
\end{equation}

\vskip\baselineskip
\noindent
, where $\beta$ represents a coefficient, \textit{s} represents a state, \textit{t} represents a day, and $\epsilon$ represents an error term.

Model 1 was the main model. The independent variable was $\ln(\textit{COVID}_{st-1} + 1)$, representing a log of the 7-day moving average of the number of COVID-19 cases, and the dependent variable was $\ln(\textit{UncivilPosts}_{st} + 1)$, denoting a log of the number of uncivil posts directed at the state governor. In this model, \textit{COVID} on a given day was assumed to affect \textit{UncivilPosts} on the next day. This time delay was introduced because the number of COVID-19 cases on a given day should generally be reported on the next day or thereafter. Moving average is introduced for \textit{COVID} because people may evaluate the number of COVID-19 cases over the last few days rather than only on the previous day.

Model 1 includes \textit{StateFE} (state fixed effects) and \textit{DayFE} (day fixed effects), which can remove the omitted variable bias caused by state-specific variables that do not vary across time and time-specific variables that do not vary across states (see Stock \& Watson, 2020)\nocite{stock2020}. 

As a control variable, Model 1 included $\ln(\textit{PCR}_{st-1} + 1)$ (a log of the 7-day moving average of the number of PCR tests). Considering that some people evaluate the number of COVID-19 cases based on the number of PCR tests and that an increase in the number of PCR tests leads to an automatic increase in the apparent number of COVID-19 cases, \textit{PCR} may affect both \textit{COVID} and \textit{UncivilPosts}; thus, it should be controlled as a confounder.

Model 1 also included $\ln(\textit{GovernorPosts}_{st} + 1)$ (a log of the 7-day moving average of the numbers of governor posts), \textit{Lockdown} (a dummy indicator for lockdown), and \textit{Mask} (a dummy indicator for strong mask requirement).
When the number of COVID-19 cases is high, governors are likely to post a lot on social media to call for attention, which may in turn lead to an automatic increase in the number of uncivil posts to governors. Moreover, when the number of COVID-19 cases is high, lockdown and strong mask requirements are likely to be implemented, which may in turn stoke people's frustration and thus increase the number of uncivil posts; in this sense, these three variables are considered mediator variables. Because the causal relationship via these mediator variables is not the present study's interest, they should be controlled.

As already mentioned, all quantitative variables were converted into natural logarithm values in Model 1.\footnote{As the minimum values of the variables were 0, the present study added 1 to them before converting them into their natural logarithmic values.} The effect size of one increase in COVID-19 cases on the number of uncivil posts might largely vary depending on the sizes of the states' populations; hence, converting them into natural logarithm values was a better option, which allowed for the estimation of the relationship by which a 1\% increase in the independent variable led to a $\beta$\% increase in the dependent variable (see Stock \& Watson, 2020)\nocite{stock2020}. 

Models 2-4 were additional constructs. In Model 2, $\ln(\textit{GovernorPosts}_{st} + 1)$ was omitted, while in Model 3, \textit{Lockdown} and \textit{Mask} were omitted. As mentioned above, Model 1 controlled for $\ln(\textit{GovernorPosts}_{st} + 1)$, \textit{Lockdown}, and \textit{Mask} because they were considered mediator variables, and the present study was not interested in the causal effect via these variables. However, it has been pointed out that when an unobserved variable U affects both the mediator and dependent variables, controlling for the former can lead to a biased causal estimation (see Elwert \& Winship, 2014; Rohrer, 2018)\nocite{Elwert2014,Rohrer2018}. As it was unknown whether such a variable U existed, the present study estimated Models 2 and 3, which did not include these mediator variables.

Model 4 additionally included $\textit{StateFE} \times \textit{Trend}$, which represented state-specific time trends. This can capture regional characteristics that linearly evolve over time.
\footnote{A time trend is a quantitative variable that takes the values 1, 2, 3, ..., and 365, for day 1, day 2, day 3, ..., and day 365, respectively. A state-specific time trend is an interaction between a state dummy and a time trend  (see Angrist \& Pischke, 2014; Carpenter, 2005; Friedberg, 1998)\nocite{angrist2014,Carpenter2005,Friedberg1998}.
}

The linear regression models were estimated with OLS, and \textit{t}-tests on the estimated $\beta_1$ were conducted at a significance level of \textit{p} = .05.

\section*{Results}
\begin{figure}[H]
  \centering
  \includegraphics[scale=0.50]{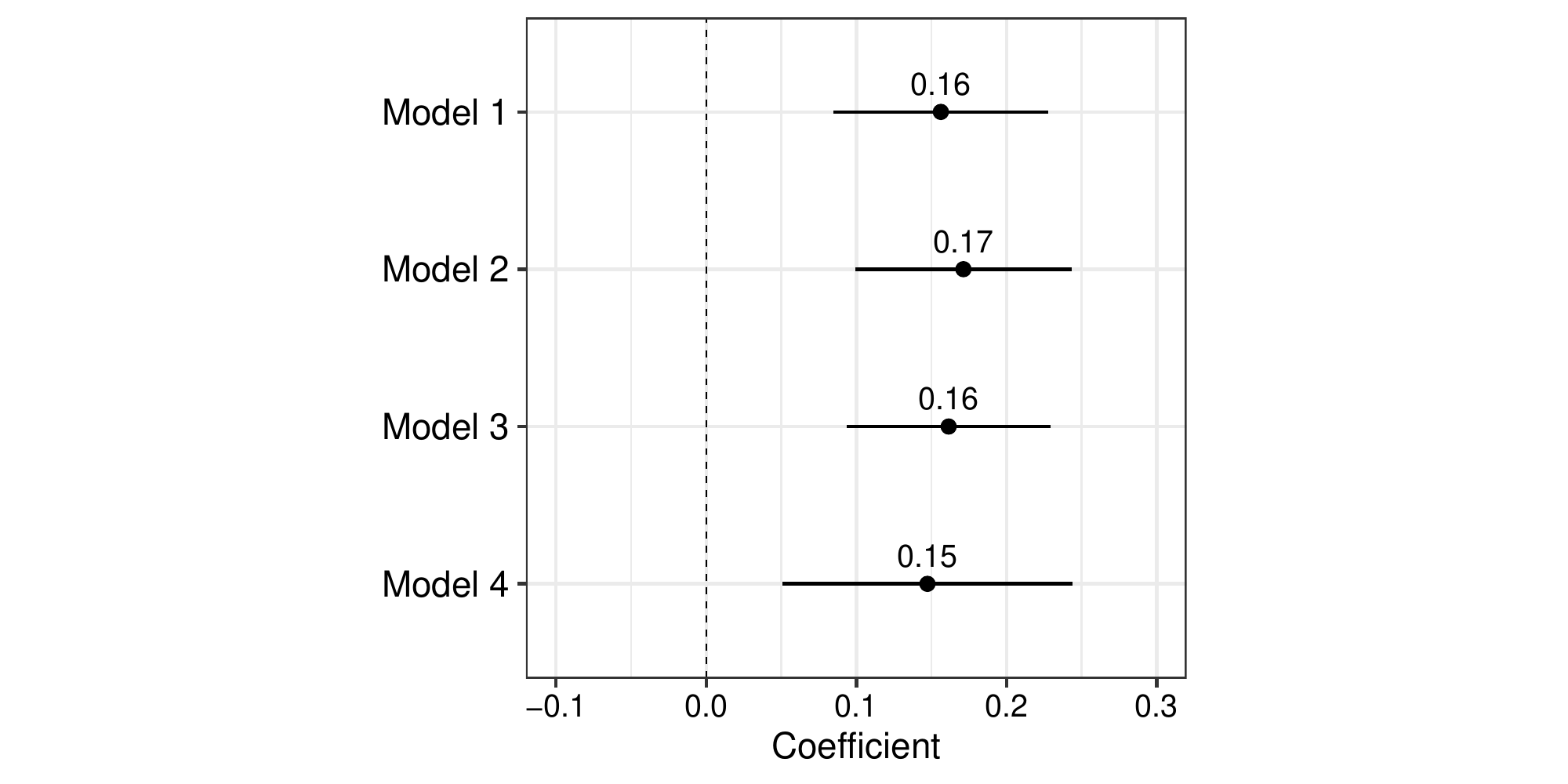}
  \vspace{-10mm}
  \caption{Main Results}
  \label{main_results_fig}
\end{figure}

Figure \ref{main_results_fig} presents the estimated $\beta_1$ for each model with 95\% confidence intervals.
The result of Model 1 shows that the estimated $\beta_1$ is 0.16, which is statistically significant (\textit{t} = 4.39, \textit{p} < .001).
This means that a 1\% increase in the 7-day moving average of the number of state-level COVID-19 cases led to a 0.16\% increase in the number of uncivil posts directed at governors. In addition, the standardized regression coefficient of the independent variable from Model 1 is .15. To evaluate the magnitude of the effect size based on the standardized regression coefficient, here the present study uses a criterion for the Pearson correlation coefficient \textit{r}. According to Funder and Ozer (2019)\nocite{Funder2019}, \textit{r} = .05 is very small, \textit{r} = .10 is small, and \textit{r} = .20 is a medium effect size. Based on this criterion, the effect size of the independent variable is small but non-negligible.
Given that past posts accumulate on social media platforms over time, this effect size can be considered not only statistically but also substantially meaningful.

Furthermore, the results of the additional Models 2-4 show that the estimated $\beta_1$ values are 0.17, 0.16, and 0.15, respectively. These estimated values are similar to the one from Model 1, and they hold statistical significance.
Therefore, the present study concludes that the hypothesis is supported.
More detailed results are presented in Table \ref{main_results_table}.

\begin{figure}[H]
  \centering
  \includegraphics[scale=0.50]{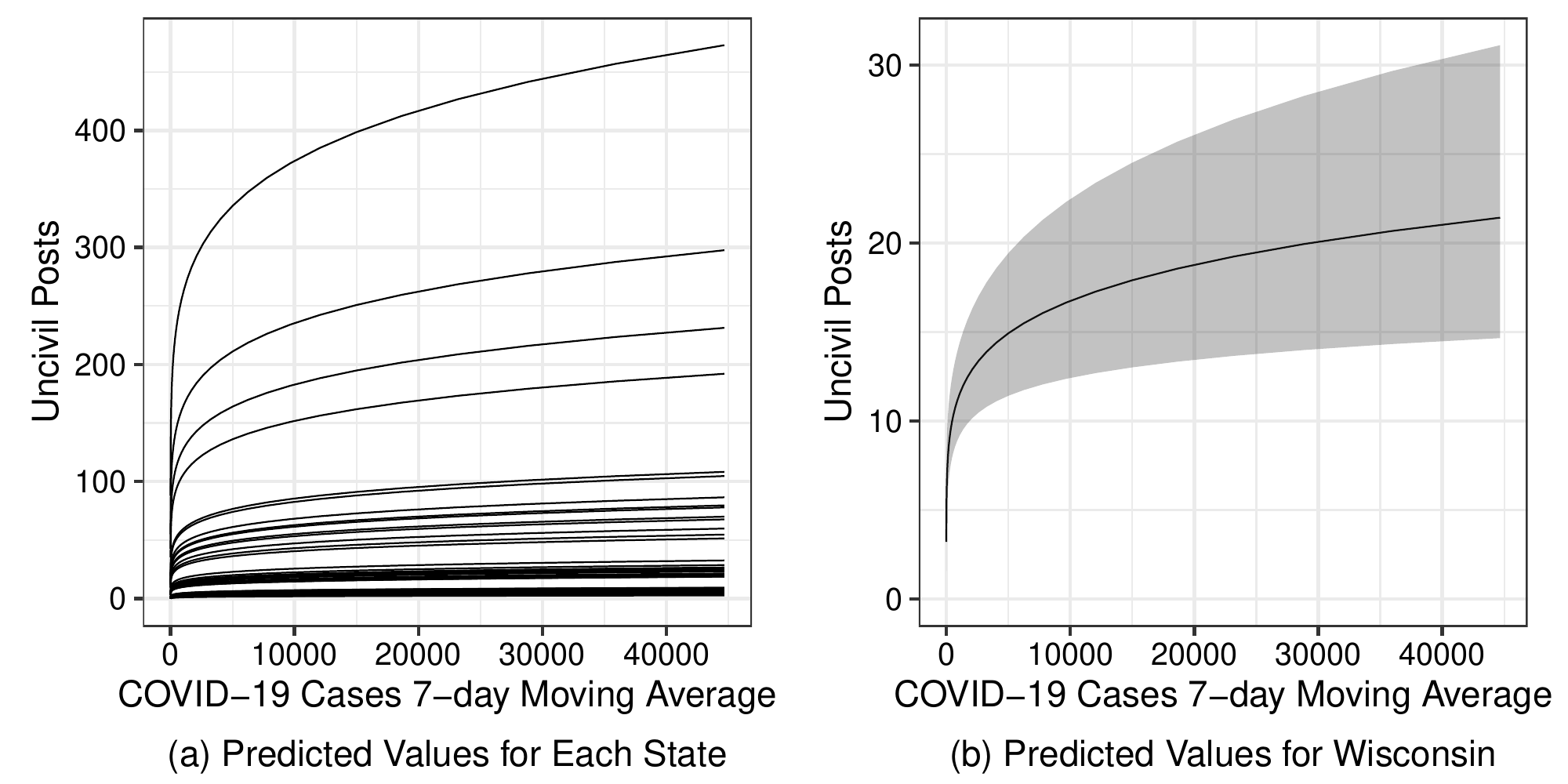}
  \vspace{-10mm}
  \caption{Predicted Values from Regression Model 1}
  \label{predicted_values}
\end{figure}

Figure \ref{predicted_values} visualizes the predicted values from regression Model 1. The left panel (a) presents the predicted values for each of the 50 states, while the right panel (b) presents the predicted values for a representative state, Wisconsin, with their 95\% confidence intervals.\footnote{Predicted values and their confidence intervals (only in panel (b)) were calculated with the predictions() function in the marginaleffects package (Arel-Bundock, 2023)\nocite{Arel-Bundock2023} in R. When predicted values were calculated, the control variables were set to their representative values. More specifically, $\ln(\textit{PCR} + 1)$ and $\ln(\textit{GovernorPosts} + 1)$ were set to their mean values. \textit{Lockdown} and \textit{Mask} were set to their mode values. \textit{DayFE} was set to January 26, 2021, because the coefficient for its dummy variable was closest to the mean value of the coefficients for the date dummy variables. For panel (b), Wisconsin was chosen as the representative state because the coefficient for its dummy variable was closest to the mean value of the coefficients for the state dummy variables.}

\vspace{3mm}

\section*{Robustness Check}
As a robustness check, the way of counting uncivil posts was modified. More specifically, if a given user sent more than one uncivil post to the governor in a day, the number of uncivil posts by that user on that day was counted as one. A user might have sent a large number of uncivil posts; this could lead to the tendencies of a few such non-general people being overly reflected in the estimations; in this case, counting uncivil posts in this modified way might appropriately address such bias. Using the new dataset, Models 1-4 were again estimated.

\begin{figure}[H]
  \centering
  \includegraphics[scale=0.50]{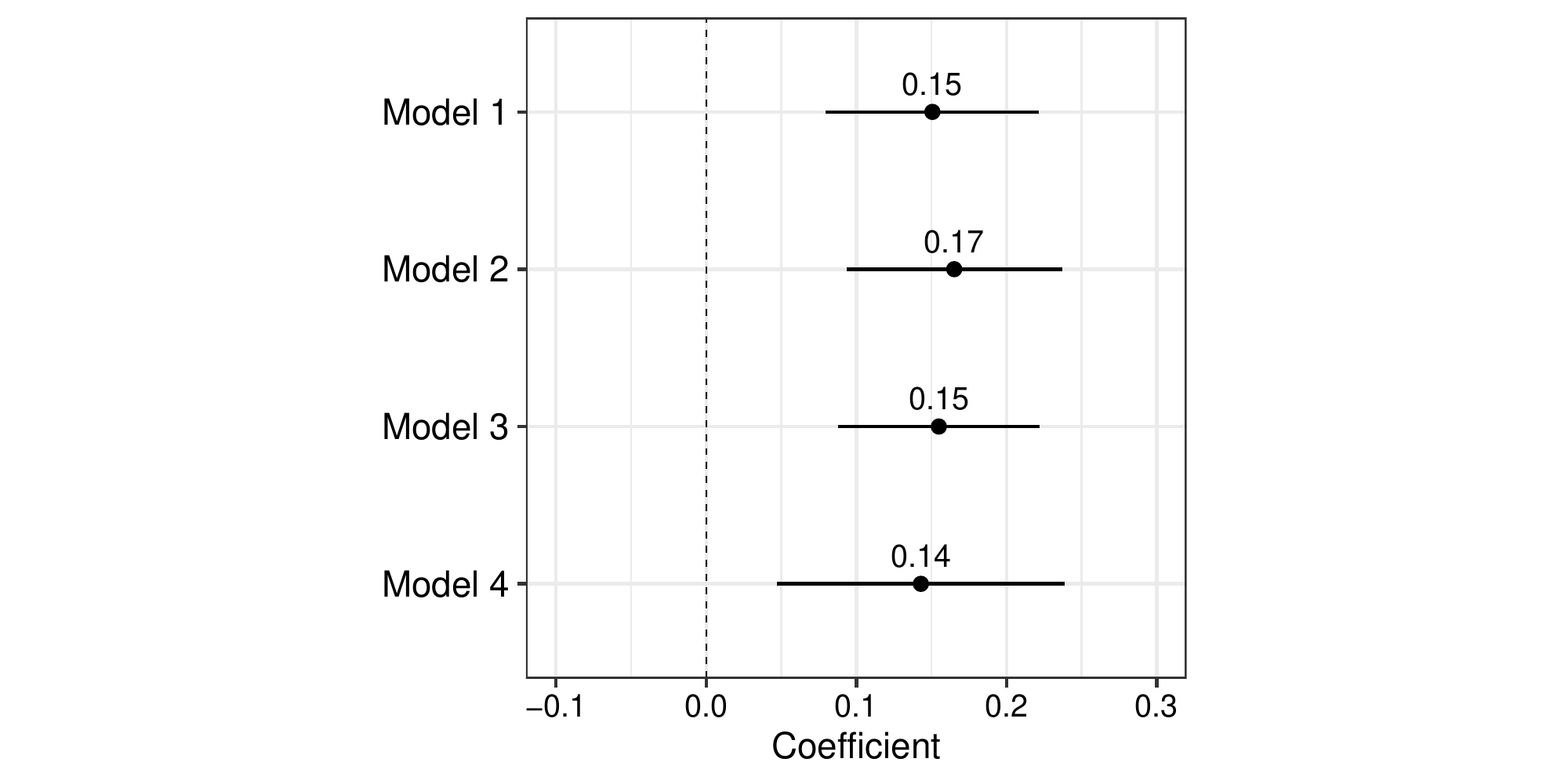}
  \vspace{-10mm}
  \caption{Results of a Robustness Check}
  \label{robustness_results_fig}
\end{figure}

Figure \ref{robustness_results_fig} shows the estimated $\beta_1$ for each model with 95\% confidence intervals. The result of Model 1 shows that the estimated $\beta_1$ is 0.15, which is statistically significant (\textit{t} = 4.27, \textit{p} < .001) and similar to the estimated value in the main results.
Furthermore, the results of Models 2-4 show that the estimated $\beta_1$ values are 0.17, 0.15, and 0.14, respectively. These estimated values are also similar to those from the main results, and are statistically significant.
Therefore, the results of the robustness check support the hypothesis.
More detailed results are given in Table \ref{robustness_results_table}.

\section*{Discussion}
The present study investigated whether the worsening of government performance led to an increase in the number of uncivil posts on X/Twitter directed at the government. The present study collected posts on X/Twitter directed at US state governors and classified them as uncivil or civil via a machine learning method. Two-way fixed effects regression models were estimated, which revealed that an increase in COVID-19 cases led to a statistically significant increase in the number of uncivil posts against state governors.

Sophisticated methodologies were employed to provide solid evidence. Adopting a neural network machine learning method enabled the automatic classification of many posts as uncivil or civil. Furthermore, the introduction of the two-way fixed effects in the regression models allowed for the removal of a large part of omitted variable bias and thus provided causal evidence.

The present study makes a significant contribution in that it reveals one side of how citizens manage to monitor and control their governments through social media. 
Non-institutionalized political participation, a tool for expressing political grievances and challenging authorities (Melo \& Stockemer, 2012; Norris, 2002)\nocite{Melo2012, Norris2002}, has existed for a long time in the history of democracy, such as through street protest demonstrations and petitions. 
With the recent development of the Internet, political expression on social media has become an established form of participation (Theocharis, 2015)\nocite{Theocharis2015}. 
As the digital native generation becomes an increasingly large part of the population over time, it would play an increasingly significant role in the mechanism of politics, coupled with the lack of costs in posting on social media.
Considering the findings of the previous studies that reveal various undesirable impacts of online political incivility on democratic societies (Hwang et al., 2018; Kim \& Kim, 2019; Popan et al., 2019)\nocite{Hwang2018, Kim2019, Popan2019}, people's use of uncivil expression in activities of monitoring and controlling governments might produce negative unintended outcomes for them.
Studies from such a perspective would become increasingly important in the future.

A second contribution is that the findings of the present study suggest the importance of elections in a democracy. When people elect a low-quality government, its performance would also be low. Low performance, according to the findings of the present study, increases uncivil behavior and worsens the environment of a discussion forum, and poor communication in a poor discussion forum would subsequently lead to increasingly low government performance. Through this process, a democratic society might fall into a negative spiral. This suggests that even one wrong choice in an election could be irreversible, and therefore, every election is crucial for maintaining a healthy democracy.

Despite these contributions, the present study has several limitations.
First, the present study focused on the US X/Twitter sphere; thus, whether the findings of the present study apply to regions outside the US or to social media platforms other than X/Twitter has not been ascertained. Hence, validating the findings of the present study in various regions and platforms is an important future task.
Second, although the data in the present study are limited to replies and mentioned posts directed at state governors, some people send uncivil posts in other ways. In addition, it is possible that people blame the worsening of COVID-19-related indicators not only on the governors but also on other actors, such as state health authorities and state legislators. Hence, a challenge for future scholars will be to analyze uncivil posts directed at these actors.
Third, the present study did not consider the presence of bots. It has been argued that bots are actively producing and spreading conspiracies and hate speech in the age of the COVID-19 pandemic (Ferrara, 2020; Uyheng \& Carley, 2020)\nocite{Ferrara2020, Uyheng2020}. Although the second robustness check might have eliminated the influence of bots to some extent, future scholars should analyze data excluding bots through automated bot detection techniques.
Because social media might become increasingly popular and play a progressively important role in society in the future, more research is needed on how people engage in democratic politics through social media communication.

\theendnotes

\section*{Funding}
This study was financially funded by the JSPS KAKENHI Grant Number 22J21515, the Research Fund from the Quantitative Methods for International Studies Program at Kobe University, and the author's own resources.

\printbibliography

\startappendix
\vspace{-1.0cm}
\begin{table}[H]
    \centering\tablesize
    \caption{Descriptive Statistics}
    \label{descriptive}
    \begin{tabularx}{125mm}{lRRRR}
    \hline
    {} & \textit{M} & \textit{SD} & \textit{Min} & \textit{Max} \\
    \hline
    Uncivil posts          &      54.18 &      269.86 &         0.00 &    26,195.00 \\
    Uncivil posts (log)    &       2.48 &        1.68 &         0.00 &        10.17 \\
    COVID-19 cases          &   1,777.70 &    3,279.55 &         0.00 &    59,712.00 \\
    COVID-19 cases (log)    &       6.44 &        1.67 &         0.00 &        11.00 \\
    PCR tests               &  20,850.28 &   31,252.62 &         0.00 &   371,033.00 \\
    PCR tests (log)         &       9.14 &        1.45 &         0.00 &        12.82 \\
    Governor posts         &       3.64 &        5.63 &         0.00 &        98.00 \\
    Governor posts (log)   &       1.12 &        0.87 &         0.00 &         4.60 \\
    Lockdown                &       0.18 &        0.38 &         0.00 &         1.00 \\
    Strong mask requirement &       0.65 &        0.48 &         0.00 &         1.00 \\
    \hline
    \end{tabularx}
\end{table}

\begin{table}[H]
\centering\tablesize
\caption{Main Results} 
\label{main_results_table}
\begin{tabularx}{160mm}{lYYYY}
  \hline
 & Model 1 & Model 2 & Model 3 & Model 4 \\ 
  \hline
COVID-19 cases 7MA (log) &  0.16 &  0.17 &  0.16 &  0.15 \\ 
   & [ 0.08, 0.23] & [ 0.10, 0.24] & [ 0.09, 0.23] & [ 0.05, 0.24] \\ 
   & \textit{SE} = 0.04 & \textit{SE} = 0.04 & \textit{SE} = 0.03 & \textit{SE} = 0.05 \\ 
   & \textit{t} =  4.39 & \textit{t} =  4.77 & \textit{t} =  4.77 & \textit{t} =  3.07 \\ 
   & \textit{p} < .001 & \textit{p} < .001 & \textit{p} < .001 & \textit{p} = .003 \\ 
  PCR tests 7MA (log) & -0.09 & -0.09 & -0.09 & -0.06 \\ 
   & [-0.19, 0.01] & [-0.18, 0.01] & [-0.19, 0.01] & [-0.19, 0.07] \\ 
   & \textit{SE} = 0.05 & \textit{SE} = 0.05 & \textit{SE} = 0.05 & \textit{SE} = 0.06 \\ 
  Governor posts 7MA (log) &  0.65 &  &  0.65 &  0.61 \\ 
   & [ 0.51, 0.78] &  & [ 0.52, 0.78] & [ 0.49, 0.73] \\ 
   & \textit{SE} = 0.07 &  & \textit{SE} = 0.07 & \textit{SE} = 0.06 \\ 
  Lockdown &  0.08 &  0.11 &  & -0.03 \\ 
   & [-0.05, 0.21] & [-0.03, 0.25] &  & [-0.14, 0.08] \\ 
   & \textit{SE} = 0.07 & \textit{SE} = 0.07 &  & \textit{SE} = 0.06 \\ 
  Strong mask requirement &  0.01 &  0.04 &  &  0.11 \\ 
   & [-0.16, 0.17] & [-0.13, 0.22] &  & [-0.02, 0.23] \\ 
   & \textit{SE} = 0.08 & \textit{SE} = 0.09 &  & \textit{SE} = 0.06 \\ 
   \hline
State fixed effect & Yes & Yes & Yes & Yes \\ 
  Day fixed effect & Yes & Yes & Yes & Yes \\ 
  Time trend & No & No & No & Yes \\ 
   \hline
\textit{N} of observations & 18,250 & 18,250 & 18,250 & 18,250 \\ 
   \hline
\end{tabularx}

\begin{tabularx}{160mm}{L}
\tablenotesize
Numbers above blankets are regression coefficients for each variable. Numbers in blankets are 95\% confidence intervals of the coefficients. Standard errors are clustered by state. 7MA means 7-day moving average. The dependent variable is the number of uncivil posts (log).\\
\end{tabularx}
\end{table}

\begin{table}[H]
\centering\tablesize
\caption{Results of a Robustness Check}
\label{robustness_results_table}
\begin{tabularx}{160mm}{lYYYY}
  \hline
 & Model 1 & Model 2 & Model 3 & Model 4 \\ 
  \hline
COVID-19 cases 7MA (log) &  0.15 &  0.17 &  0.15 &  0.14 \\ 
   & [ 0.08, 0.22] & [ 0.09, 0.24] & [ 0.09, 0.22] & [ 0.05, 0.24] \\ 
   & \textit{SE} = 0.04 & \textit{SE} = 0.04 & \textit{SE} = 0.03 & \textit{SE} = 0.05 \\ 
   & \textit{t} =  4.27 & \textit{t} =  4.63 & \textit{t} =  4.64 & \textit{t} =  3.00 \\ 
   & \textit{p} < .001 & \textit{p} < .001 & \textit{p} < .001 & \textit{p} = .004 \\ 
  PCR tests 7MA (log) & -0.08 & -0.08 & -0.08 & -0.06 \\ 
   & [-0.18, 0.02] & [-0.17, 0.01] & [-0.18, 0.01] & [-0.19, 0.07] \\ 
   & \textit{SE} = 0.05 & \textit{SE} = 0.05 & \textit{SE} = 0.05 & \textit{SE} = 0.06 \\ 
  Governor posts 7MA (log) &  0.63 &  &  0.63 &  0.59 \\ 
   & [ 0.50, 0.76] &  & [ 0.50, 0.76] & [ 0.48, 0.70] \\ 
   & \textit{SE} = 0.06 &  & \textit{SE} = 0.06 & \textit{SE} = 0.06 \\ 
  Lockdown &  0.07 &  0.10 &  & -0.04 \\ 
   & [-0.06, 0.20] & [-0.04, 0.23] &  & [-0.15, 0.07] \\ 
   & \textit{SE} = 0.06 & \textit{SE} = 0.07 &  & \textit{SE} = 0.06 \\ 
  Strong mask requirement &  0.00 &  0.04 &  &  0.11 \\ 
   & [-0.17, 0.17] & [-0.13, 0.22] &  & [-0.02, 0.23] \\ 
   & \textit{SE} = 0.08 & \textit{SE} = 0.09 &  & \textit{SE} = 0.06 \\ 
   \hline
State fixed effect & Yes & Yes & Yes & Yes \\ 
  Day fixed effect & Yes & Yes & Yes & Yes \\ 
  Time trend & No & No & No & Yes \\ 
   \hline
\textit{N} of observations & 18,250 & 18,250 & 18,250 & 18,250 \\ 
   \hline
\end{tabularx}

\begin{tabularx}{160mm}{L}
\tablenotesize
Numbers above blankets are regression coefficients for each variable. Numbers in blankets are 95\% confidence intervals of the coefficients. Standard errors are clustered by state. 7MA means 7-day moving average. The dependent variable is the number of uncivil posts (log).\\
\end{tabularx}
\end{table}

\end{document}